\documentclass[12pt]{article}
\def\1{\'{\i}}
\def\3{\ss}

\usepackage{amsfonts}
\usepackage{graphicx}

\title{Computer Simulation of Convective Plasma Cells}

\author{Rodrigo~Carboni and Francisco~Frutos-Alfaro \\ 
{\small Space Research Center (CINESPA) and School of Physics} \\
{\small University of Costa Rica, San Jos\'e, Costa Rica} \\
{\small Emails: rcarboni@cariari.ucr.ac.cr, frutos@fisica.ucr.ac.cr}}

\date{\today}

\begin{document}

\maketitle

\pagestyle{headings}

\begin{abstract}
\noindent
Computer simulations of plasmas are relevant nowadays, because it helps us 
understand physical processes taking place in the sun and other stellar 
objects. We developed a program called {\tt PCell} which is intended for 
displaying the evolution of the magnetic field in a 2D convective plasma cell 
with perfect conducting walls for different stationary plasma velocity fields. 
Applications of this program are presented. This software works interactively 
with the mouse and the users can create their own movies in {\tt MPEG} format. 
The programs were written in {\tt Fortran} and {\tt C}. 
There are two versions of the program ({\tt GNUPLOT} and {\tt OpenGL}). 
{\tt GNUPLOT} and {\tt OpenGL} are used to display the simulation.
\end{abstract}

\section{Introduction}

\noindent
The huge advance in computer technology makes it possible to simulate and 
visualize complex physical phenomena taking place in stellar objects, 
{\it e.g.}, the convective plasma cells on the sun. 

\noindent
To understand the dynamo mechanism, most researchers consider cosmic plasma 
with a stationary motion, which leads to an induction problem. The aim is to 
find stationary states as solutions of the induction equation. 
Elsasser \cite{Elsasser}, Weiss \cite{Weiss} and Parker \cite{Parker} performed 
the first two-dimensional simulations using symmetric velocity fields. 

\noindent
Simulations by Weiss and Galloway include dynamical effects and are generalized 
to three dimensions for the kinetic case 
(Weiss \cite{Weiss}, Galloway {\it et al.} \cite{Galloway}, 
Galloway and Weiss \cite{Galloway2}). 
In recent years, dynamo models have been improved by considering fully
dynamical solutions of the induction equation taking into account the coupled 
mass, momentum and energy relations for the plasma. Recently, Mininni et al. 
\cite{Minini} have considered the Hall current into the dynamo model.

\noindent
We can simulate convection cells by choosing convective velocity fields, which 
in turn help us understand the behavior of granules, mesogranules in the 
photosphere, supergranules in the photosphere and chromosphere, and laboratory 
plasmas. We also can investigate the not-well-understood phenomena of
reconnection in this way \cite{Carboni}. 

\noindent
The {\tt PCell} program \cite{Carboni2} helps visualize the magnetic field 
evolution in different convective plasmas. This program runs on {\tt Linux} or 
{\tt Unix}. The program can be downloaded from

{\tt http://cinespa.ucr.ac.cr/software/xpcell/} 

\noindent
The capability to easily create movies in {\tt MPEG} format is one of the main 
features of {\tt PCell}.

\section{The Induction Equation}

\noindent
Maxwell's Equations determine electromagnetic fields behavior in a cosmic 
fluid

\begin{eqnarray}
\label{maxwell}
{\epsilon} {\bf \nabla} \cdot {\bf E} & = & {\rho} \\  
{\bf \nabla} \times {\bf E} & = & - {{\partial {\bf B}}\over{\partial t}} , \\
{\bf \nabla} \cdot {\bf B} & = & 0 , \\ 
{\bf \nabla} \times {\bf B} & = & \mu {\bf J} 
+ \mu \epsilon {\partial {\bf E} \over \partial t} , 
\end{eqnarray}

\noindent
where $ \rho $ is the charged particles density. The effects produced by the 
temperature gradient and charged particles density fluctuations had been 
neglected.

\noindent
The general expression for the current density $ {\bf J} $ in an isotropic 
homogeneous medium is:

\begin{equation}
\label{def_j}
{\bf J} = \sigma {\bf E} + \sigma {\bf v} \times {\bf B} + \rho {\bf v}  
\end{equation} 

\noindent
where the right-hand-side terms are the conduction, induction and convection 
currents respectively. The conductivity $ \sigma $ is considered constant in 
the whole plasma and $ {\bf v} $ is the velocity field that describes the 
plasma motion.

\noindent
We can simplify the above equations by comparing the orders of magnitude 
of the quantities involved. We represent orders of magnitude with square 
brackets. % ($ \[ \, \] $).

\noindent
In cosmic plasmas the velocity of the charged particles 
(mechanical velocities) are much slower the electromagnetic field velocity 
(speed of light). Therefore, we have:

\begin{equation}
{\left[ {v \over c} \right]} = [ \beta ] \ll 1 ,    
\end{equation}

\noindent
which implies that the orders higher than $ \beta $ can be neglected 
(non relativistic plasma).

\noindent
From the Faraday equation (the second equation of \ref{maxwell}), we can 
obtain an quantity with dimensions of velocity:

\begin{equation}
[ E ] = [ v_{el} B ] ,
\end{equation}

\noindent
which is the velocity associated to the electromagnetic processes and 
satisfies the condition:

\begin{equation}
\label{comp_vel}
[ v_{el} ] \le [ v ] .  
\end{equation} 

\noindent
The last two equations combined give us an estimation of the rate of the 
electric energy to the magnetic energy:

\begin{equation}
{\left[ {\epsilon E^2 \over \mu^{-1} B^2} \right] } = 
\left[ {E^2 \over c^2 B^2 } \right] \le [ \beta^2 ] . 
\end{equation}

\noindent
If equation (\ref{comp_vel}) is fullfil, the electric component 
$ \rho {\bf E} $ of the magnetic force that exerts the magnetic field over the 
plasma is negligible compared with the magnetic component 
$ {\bf J} \times {\bf B} $.

\noindent
The displacement current in the Ampere-Maxwell equation (the last equation 
of \ref{maxwell}) is negligible when is compared to the conduction current. 
The rate of both is given by $\gamma $:

\begin{equation}
\label{def_gamma}
\gamma = \left[ {\omega_{el} \epsilon \over \sigma} \right] ,    
\end{equation}

\noindent
where $ \omega_{el} $ is the electromagnetic frequency. Taking $ L $ as a 
characteristic length of the electromagnetic and mechanical 
phenomena \cite{Elsasser}, we have %(Elsasser, 1956)

\begin{equation}
[ v ] = [ L \omega ] ,
\end{equation} 

\noindent
and from (\ref{def_gamma}) we obtain

\begin{equation}
[ \omega_{el} ] \le [ \omega ] .
\end{equation} 

\noindent
For the Earth's core $ \gamma \le 10^{-18} $ and for stars 
$ \gamma \ll 1$ \cite{Dolginov}. % (Dolginov, 1988).

\noindent
The rate of the convection current to the conduction current has the same 
value $ \gamma $. From the Gauss law for the electric field (first equation 
of \ref{maxwell}) we have that $ [ \rho ] = [ {\epsilon E / L}] $, therefore

\begin{equation}
\label{conv/conv}
\left[ {\rho v \over \sigma E} \right] = 
\left[ {{\epsilon v} \over {\sigma L}} \right] = [ \gamma ] .
\end{equation} 

\noindent
According to equations (\ref{def_gamma}) and (\ref{conv/conv}), equation 
(\ref{def_j}) and the last equation of (\ref{maxwell}) simplify to 

\begin{equation}
{\bf J} = \sigma {\bf E} + \sigma {\bf v} \times {\bf B} ,
\end{equation}

\noindent
and 

\begin{equation}
{\bf \nabla} \times {\bf B} = \mu {\bf J} .
\end{equation}
respectively.

\noindent
Combining the latter two equations we obtain

\begin{equation}
\label{casi_ecua_ind}
{\bf \nabla} \times {\bf B} = \mu {\bf J} = \mu \sigma {\bf E} 
+ \mu \sigma {\bf v} \times {\bf B} .
\end{equation}

\noindent
To eliminate $ {\bf E} $, we apply the curl to equation (\ref{casi_ecua_ind}) 
and make the substitution of Faraday's law (\ref{maxwell}), yielding

\begin{equation}
\mu \sigma {\partial {\bf B} \over \partial t} = 
{\mu \sigma {\bf \nabla} \times ({\bf v} \times {\bf B} )} 
- {{\bf \nabla} \times {\bf \nabla} \times {\bf B}} .
\end{equation}

\noindent
Finally, if we change the second term of the right side with the help of the 
well known identity and use the magnetic Gauss law (third equation of 
\ref{maxwell}) we arrive to the induction equation

\begin{equation}\label{ecu_ind}
{\partial {\bf B} \over \partial t} = 
{{\bf \nabla} \times ({\bf v} \times {\bf B})} 
+ {\eta {\bf \nabla}^2 {\bf B}} ,
\end{equation} 

\noindent
where $ \eta = {1 / \mu \sigma} $ is the magnetic viscosity.

\section{Behavior of the Induction Equation}

The rate of the first term on the right side of the induction equation to the 
second one is given by $ R_m = {L v / \eta} $, where the adimensional 
quantity is called the magnetic Reynolds number, in analogy to the Reynolds 
number for non conducting fluids. The bigger the plasma characteristic lengths 
is, the bigger the magnetic Reynolds number.

\noindent
If the first term is much bigger than the second, equation (\ref{ecu_ind}) 
can be written as

\begin{equation}
{{\partial {\bf B}} \over {\partial t}} = \eta {\bf \nabla}^2 {\bf B}.
\end{equation} 

\noindent
This is the diffusion equation, which describes the magnetic field decay in a 
diffusion characteristic time
 
\begin{equation}
{\tau_\eta} = {L^2 \over 4 \pi^2 \eta}
\end{equation}

\noindent
for a plasma with spherical symmetry. It is of the order of one second for a 
one centimeter radii copper sphere, $ 10^4 $ years for the Earth's nucleus 
and $ 10^{10} $ years for the Sun.

\noindent
The order of the magnetic Reynolds number can be written as

\begin{equation}
[ R_m ] = \left[ {\omega \over \omega_{el}} \right] 
= \left[ {\tau_\eta \over\tau_0} \right] 
\end{equation}

\noindent
where $ \tau_0 $ is the time associated to plasma mechanical motion. 
Therefore, for times much slower than the diffusion time the equation 
simplifies to

\begin{equation}
{{\partial {\bf B}} \over {\partial t}}  =  
{\bf \nabla} \times ({\bf v} \times {\bf B})
\end{equation}

\noindent
which states that the magnetic flux trough any closed curve that moves with 
the local velocity of the plasma remains constant in time, i.e., the magnetic 
field lines are dragged by the fluid ({\it frozen field lines}).  

\noindent
When $ R_m \gg 1 $ the transport of the field lines by the plasma dominates 
over the diffusion, but if $ R_m \ll 1 $ the field decays very fast and the 
dynamo effect cannot take place. The behavior generated from the interplay of 
both terms for magnetic Reynolds number values between these limits is very 
interesting and to explore it is the aim of this work.

\section{The Vector Potential Function}

\noindent
The induction equation (\ref{ecu_ind}) can be simplified for the two 
dimensional case if it is written as a function of the vector potential. 
We take the magnetic field and the velocity field limited to the $ x - y $ 
plane, then the magnetic field is obtained from the one component vector 
potential $ {\bf A}  =  A {\bf k} $ as follows  

\begin{equation}
{\bf B} = {\bf \nabla} \times {\bf A} 
= \left({\partial A \over \partial y}, \, - {\partial A \over \partial x} , 
\, 0 \right) ,
\end{equation} 

\noindent
after the substitution, equation (\ref{ecu_ind}) yields

\begin{equation}
{{\partial A} \over {\partial t}} = - {\bf u} \cdot  {\bf \nabla} A 
+ \eta {\bf \nabla}^2  A .
\end{equation}

\noindent
We define the position and velocity variables as function of the 
characteristic parameters (maximum velocity $ U $ and maximum length $ L $) as 
follows

\begin{equation}
u = u^\prime U , \qquad x = x^\prime L
\end{equation}

\noindent
with the analogous definition for the $ y $ coordinate. The spatial 
derivatives are given by 

\begin{equation}
{{\partial} \over {\partial x}} = {{1} \over {L}}  
{{\partial}\over {\partial x^\prime}}
\end{equation} 

\noindent
and 
\begin{equation}
{{\partial^2}  \over  {\partial x^2 }}  =  { 1  \over { L^2 }} 
{{\partial^2} \over { \partial x^{ \prime^2 }}} .
\end{equation}

\noindent
After the substitution of these relations and defining a the characteristic 
time of the mechanical motion $ \tau_0 = {{L} / {U}} $. For instance, 
if we take following velocity values for granules, mesogranules and 
supergranules in the sun, \\ 
$ V = 900, \, 60 \, {\rm and} \, 400  \, {\rm m/s} $ 
and $ L = 1.4 \times 10^3, \, 7 \times 10^3 \, {\rm and} \, 
3 \times 10^5 \, {\rm km} $, respectively, yield 
$ \tau_0 = 26 \, {\rm min}, \, 1.35 \, {\rm and} \, 0.87 \, {\rm days} $ 
(Foukal \cite{Foukal}, Sturrock {\it et al}. \cite{Sturrock}). 
This time $ \tau_0 $ measures the time it takes the plasma to go from the 
bottom to the top of the cell. With $ t = {t^\prime} \tau $, we obtain 

\begin{equation}
\label{ecua_ind_pot}
{{\partial A} \over {\partial t}} = {\bf u} \cdot {\bf \nabla} A 
+ {{1} \over { R_m }} {{\bf \nabla}^2} A ,
\end{equation}

\noindent
where the primes have been removed for clarity.

\noindent
This is the equation we solve under the kinematic condition, i.e., there is 
no reaction of the magnetic field on the plasma, leaving the velocity field 
time independent. This approach is valid if the magnetic energy is small 
compared with the kinetic energy of the plasma, that is
\begin{equation}
{B^2 \over 8 \pi \mu} \ll {1 \over 2} \rho v^2 .
\end{equation}

\section{The Visualization Program}

\subsection{Description of the Program}

\noindent
Equation (\ref{ecua_ind_pot}) is solved using a fourth order difference schema 
in a two dimensional cell with perfect conducting upper and lower walls 
(the magnetic field lines remain always tide to them) and periodic conditions 
at the lateral walls, i. e., each cell is surrounded by similar cells. 

\noindent
The velocity field is taken to be incompressible, which allow us to define a 
stream function from which it can be obtained. we chose the following stream 
function \cite{Weiss} (see Figure 2)

\begin{eqnarray} 
\psi & = & - {1 \over 4 \pi} 
\left(4 (1 - m) \left( x - {1\over2} \right)^2 - m \right) \nonumber \\
& \times & \left(1 - 4 \left(y - {1 \over 2} \right)^2 \right)^4 
\cos\pi \left(x -{1\over2} \right) 
\end{eqnarray} 

%( Weiss, 1966 )
\noindent
with $ (0 \leq x \leq 1 , \, 0 \leq y \leq 1) $, $ m $ is an adjustable 
parameter $ (0 \leq m \leq 1) $ that allow us to select different velocity 
fields. 

\noindent
When $ m = 1 $ it describe a single eddy \cite{Weiss}, as $ m $ decreases 
two new symmetrical eddies emerge from each side compressing the original 
eddie, at one time there are three eddies and as $ m $ get closer to zero, the 
central eddie disappear and remain just two eddies rotating in opposite 
directions.

\subsection{The Visualization Program}

\noindent
The program called {\tt pcell} is compressed in the {\tt pcell.tar.gz} file. 
The program was originally written in {\tt Fortran}. We translated them into 
{\tt C} with the help of {\tt F2C}. The {\tt XFORMS} library, which is used 
to design the control panel is also required. The data generated by 
the program is processed by {\tt GNUPLOT} ({\tt PCell}) or {\tt OpenGL} 
({\tt XPCell}) to produce the simulation.

\noindent
When the program starts, it creates a window: the {\it PCell Control Panel}. 
The user can control all items on it by clicking.

\subsection{The Control Panel} 

\noindent
The parameters are adjusted interactively by the user with the mouse. The 
control panel has the following items (see Figure 1):  

\begin{itemize}

\item The velocity field parameter selector: The user chooses the velocity 
field selecting a value of the $m$ parameter, $ (0 \leq m \leq 1) $.

\item The magnetic Reynolds number selector: The user selects the magnetic 
Reynolds number, $ (0 < R_m \leq 1000) $. The values of $ U $ and $ L $ are 
chosen equal to one.

\item The time window: Displays the running time in units of 
$ \tau_0  = {{L} / {U}} $.

\item The maximum running time selector: The time at which the program stops 
running can be chosen with this button.

This items can be selected in any order but they should be selected before any 
other item of the program.

\item The Draw button: This button starts the program that calculates the 
magnetic field at each time interval (these data is stored in the files 
{\tt pcell1.dat}, {\tt pcell2.dat} and so on) and after a short time, when the 
generation of data is finished, the display window opens (see Figure 3). 

\item The display window: The evolution of the magnetic field lines are 
displayed in this window. When a given velocity field is selected it is shown 
here too. 

\item The Stream button: Displays the mechanical motion of plasma 
(see Figure 2).

\item The Movie button: The users can create their own movies and with 
{\tt XANIM} or a {\tt MPEG} player they can display the movies. The name of the 
created movie is {\tt convection.mpg}.

\item The Help button: Gives the user a program guide.

\item The Exit button: To leave the session. If the user wants to remove all 
{\tt pcell*.dat} then type {\tt rm -rf pcell*.dat} on the prompt.

%The program asks if the user 
%wants to keep or discard the generated files {\tt pcell}{\it i}{\tt .dat} 
%(corresponding each one to a frame of the movie) before exiting. 

\end{itemize} 

\section{Applications}

\noindent
There are some interesting applications that the user can explore. Among these 
are:

\begin{itemize}
\item The evolution of the magnetic field lines as function of the magnetic 
Reynolds number.

\item The mechanism of magnetic field dissipation and the reconnection 
phenomena.
 
\item The evolution of the averaged magnetic density as function of the 
magnetic Reynolds number.  

\item The maximum averaged magnetic density as function of the magnetic 
Reynolds number.

\item The stationary state of the averaged magnetic density and the way it is 
reached as function of the magnetic Reynolds number.

\item The time it takes to reach the maximum and the stationary state of the 
averaged magnetic density as function of the magnetic Reynolds number.
\end{itemize}

\subsection{Evolution of the magnetic field}

\noindent
The stream function and the magnetic field twist evolution can be visualized 
easily with PCell. In Figure 2, an illustration of this evolution can be seen. 
Reconnection of the magnetic field appears in Figure 2b.

\subsection{Analysis of the magnetic field evolution}

\noindent
For high Reynolds numbers, the convection generates a magnetic field component 
parallel to the plasma motion. The field begins to amplify itself, becoming
stronger at the lateral borders of the cell (see Fig. 3). As this happens the 
spatial scale, where the variation of the field occurs, decreases linearly 
producing an increase of the diffusion. When this resistive term becomes of the
same order as the convective term, the amplification of the field stops. 
Therefore, the averaged magnetic density reaches a maximum (see Figure 4). 
The twisting of the field produces regions of high magnetic density at the same
time that the diffusive term becomes bigger and reconnection of the magnetic 
filed lines occur expelling the field from the central region of the eddy 
reaching the magnetic density a stationary value (see Figure 2f). 
This stationary value is independent of the velocity field, but the way it is 
reached depends on the velocity field (see Figure 4). 

\noindent
For velocity fields dominated by one eddy, the magnetic energy in the central 
part of the cell is very small compared with the density at the borders. As the
field is expelled from the central part by diffusion, the magnetic density does 
not change much so the stationary value comes from the field located at the
borders. For velocity fields with a strong influence of lateral eddies 
an additional region of high magnetic density appears in the middle of the 
cell, where the central eddy is located (see Fig. 3a - d). The intensity of
the field is of the same order as the one at the borders; the twisting of 
the field as it disappears on both sides of the central region of high density 
makes the average magnetic density oscillate until it reaches the stationary
value. When the velocity fields are practically two eddies (small $ m $), 
the magnetic field at the borders is not very strong and the generation of 
the central high density zone helps to reach the maximum value which is the
same as the stationary one (see Figure 4). 

\noindent
It is important that the average lifetime of the granules and mesogranules is 
very small compared to the characteristic time 
($ \sim 0.31 \, {\rm and} \, 0.07 \tau_0 $, respectively). 
Therefore for these motions, the behavior shown would not evolve because 
the structures have disappeared long before. On the other hand, in 
supergranules ($ \sim 23 \tau_0 $) the stationary state is reached.

\noindent
According to the above description for single eddies, a simple expression for 
the maximum averaged energy density and the Reynolds number can be obtained as
$ B^2_{max} \simeq R^{2/3} B^2 $, where $ B_0 $ is the initial averaged magnetic 
field (Weiss \cite{Weiss}, Carboni \cite{Carboni}). In Figure 5a, it can be
seen that this behavior is still valid for plasmas with interacting eddies. 
The Reynolds number exponents obtained for the values of 
$ m = 1.0, \, 0.7, \, 0.3 \, {\rm and} \, 0.1 $ are, respectively, 
$ 0.59, \, 0.55, \, 0.54 \, {\rm and} \, 0.67 $. In the same way, a relation 
between the stationary and averaged energy density is obtained 
(Weiss \cite{Weiss}, Carboni \cite{Carboni}) as $ B^2_{st} \simeq R^{1/2} B^2 $. 
Figure 5b shows the result of the simulations which agree very well with 
this relation even for interacting eddies. The Reynolds number exponents 
obtained in this case are $ 0.42, \, 0.42, \, 0.43 \, {\rm and} \, 0.48 $ 
in the same order as above.

\section{Conclusions and Future Work}

\noindent
The kinematical dynamo model presented here produces intensification of 
the magnetic field by the induction of plasma trying to move across field lines 
in regions of small eddies between big convective zones in addition to 
the usual one accumulated on the regions between cells. Although this model 
does not include the feedback of the $ {\bf J} \times {\bf B} $ force on 
the inducting motion and has no rotation profile, the generation of a toroidal 
field is clear. Similar oscillatory behavior of the magnetic field for bands 
of asymmetrical eddies has been found for groups of four-cell convections by 
Zegeling \cite{Zegeling} working on the same principle described here.

\noindent
The model could be implemented to solve the full dynamo problem, which involves 
the simultaneous solution of the induction equation along with the equations 
of motion, continuity and thermodynamics. The inclusion of the 
$ {\bf J} \times {\bf B} $  term produces freezing of the magnetic field to 
the electron flow instead of the bulk velocity field 
(Mininni {\it et al}. \cite{Minini}).

\noindent
Implementations of the numerical methods will allow us to expand the study, 
including the effects of nonlinearity and chaotic motion at Reynolds numbers 
typical of astrophysical problems where self-organization emerges 
(Chang {\it et al}. \cite{Chang}, Valdivia {et al}. \cite{Valdivia}).
New approaches have been developed to calculate, in a more efficient way, 
convective cells. One strategy is the adaptive grid method which, based 
on coordinate transformations between physical and computational coordinates, 
automatically tracks and spatially resolves nonlinear structures 
(Zegeling and Keppens \cite{Zegeling2}).

\bigskip
\noindent
{\it Future Work}. The program can be improved in the following way:

\begin{itemize}

\item Including more velocity fields.

\item Adding the magnetic density averaged over the cell, represented as 
function of time.

\item Expanding to three dimensional cells and using other shapes like 
hexagonal cells (this shape appears as stable patterns in some fluids).

\item Considering mechanical-electromagnetic interaction between the plas\-ma 
and the field.

\item Exploring more complex behaviors such as Chaos.

\item Using the more advance program {\tt XANIM} to produce a better animation.
\end{itemize}

\section{Acknowledgment}

\noindent
The authors would like to thank Dr.~Jorge~P\'aez for his helpful comments.

\begin{figure} %[p]
% ht = here, b = bottom, t = top, p = page of float
\label{general}
\centering
\includegraphics[width=6.5cm]{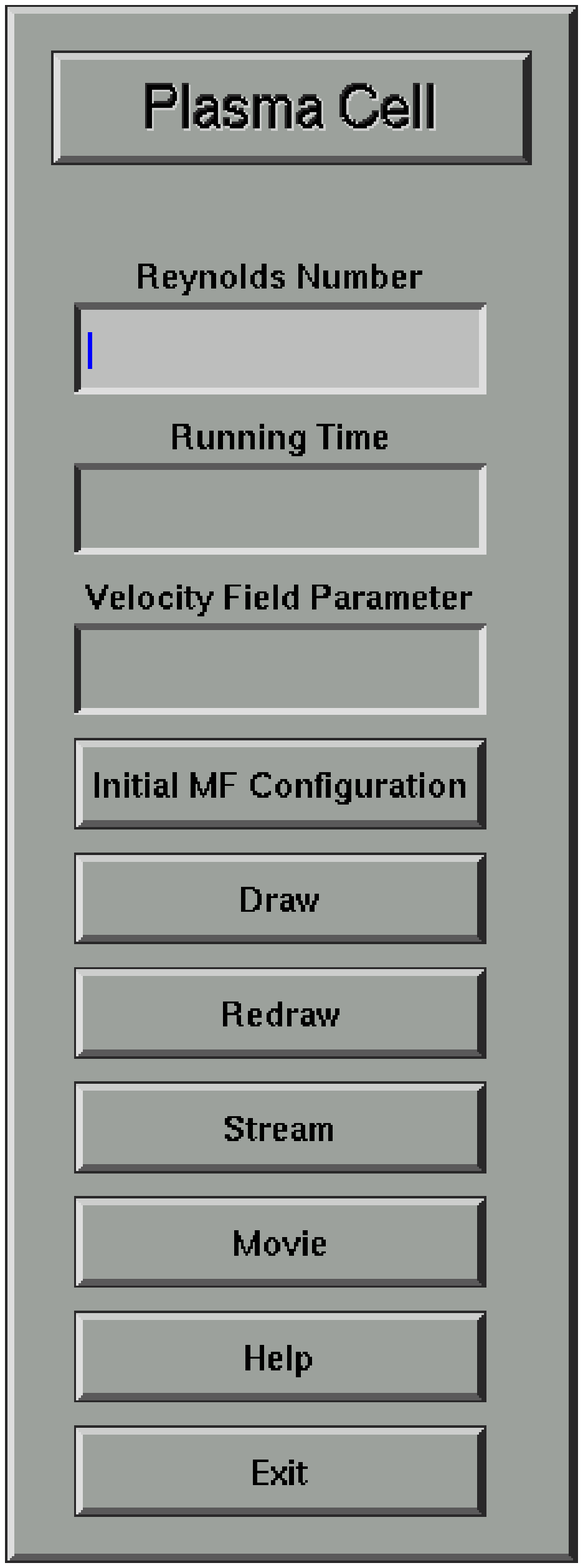}
\caption[PCell Control Panel]{PCell Control Panel}
\end{figure}

\begin{figure}[hbt] %[p]
\centerline{\unitlength1cm
\begin{picture}(15,12)
\put(4,10){\makebox(0,0){
\includegraphics[width=6.5cm,angle=90.0]{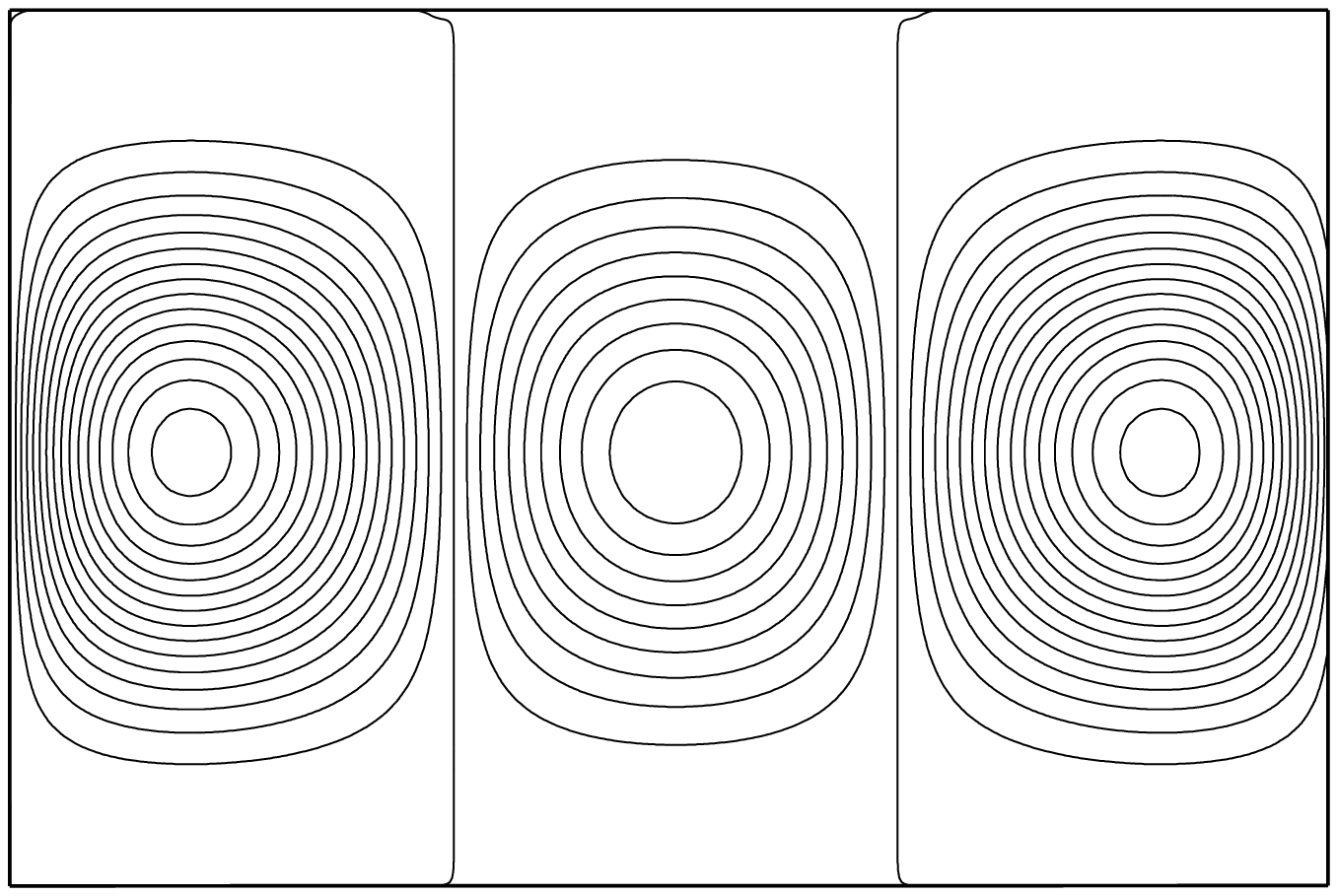}}}
\put(11,10){\makebox(0,0){
\includegraphics[width=6.5cm,angle=90.0]{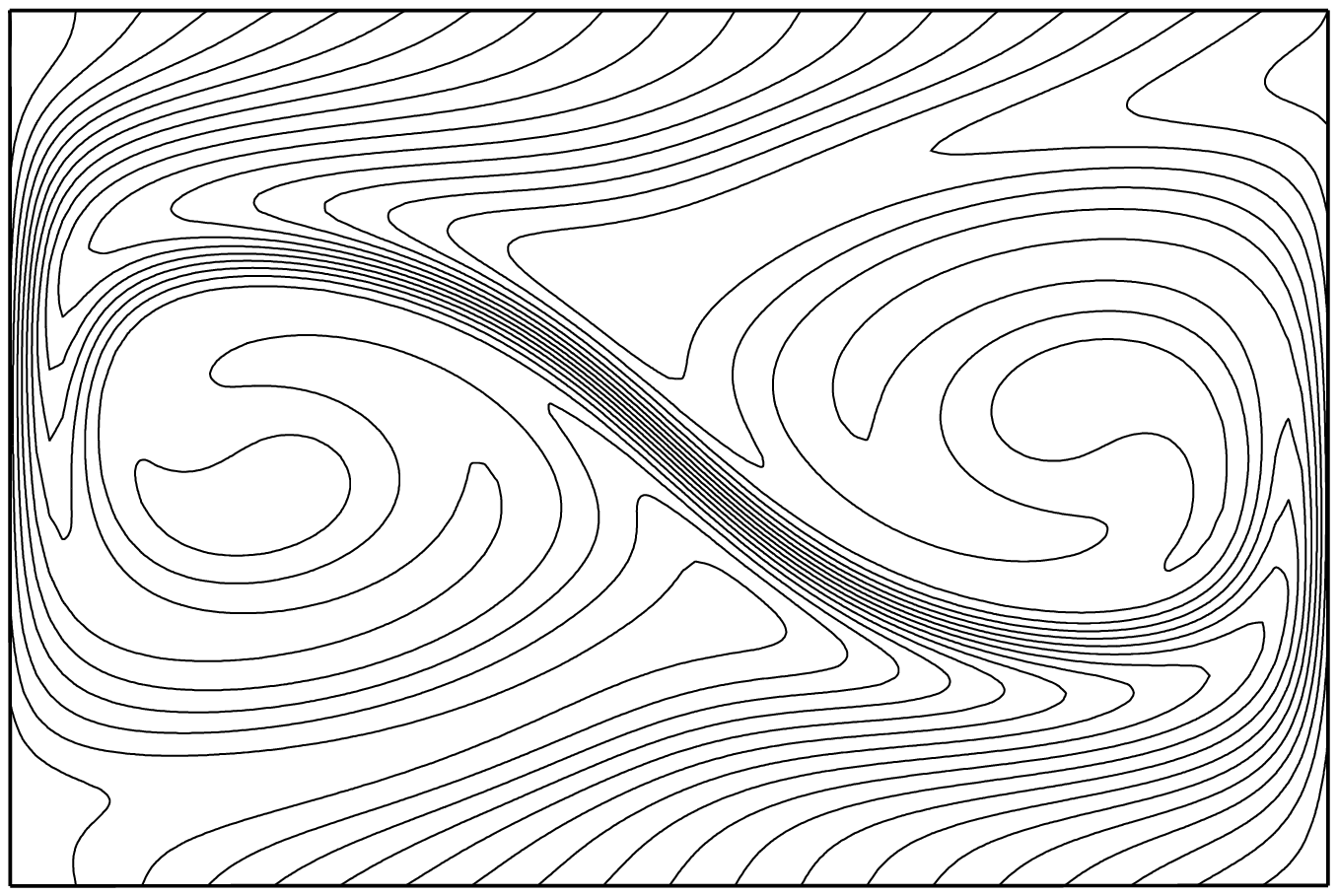}}}
\put(4,6){\makebox(0,0){
\includegraphics[width=6.5cm,angle=90.0]{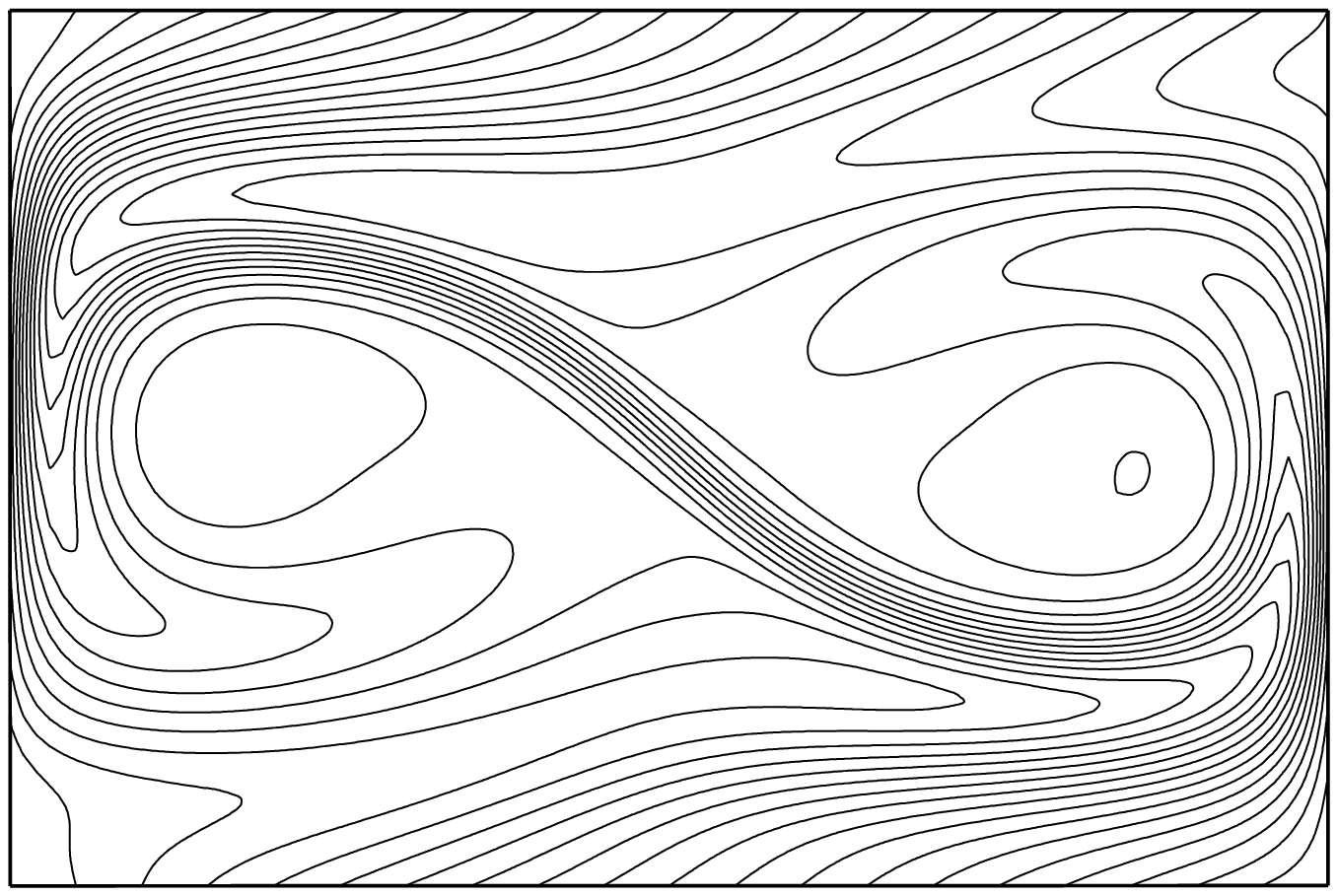}}}
\put(11,6){\makebox(0,0){
\includegraphics[width=6.5cm,angle=90.0]{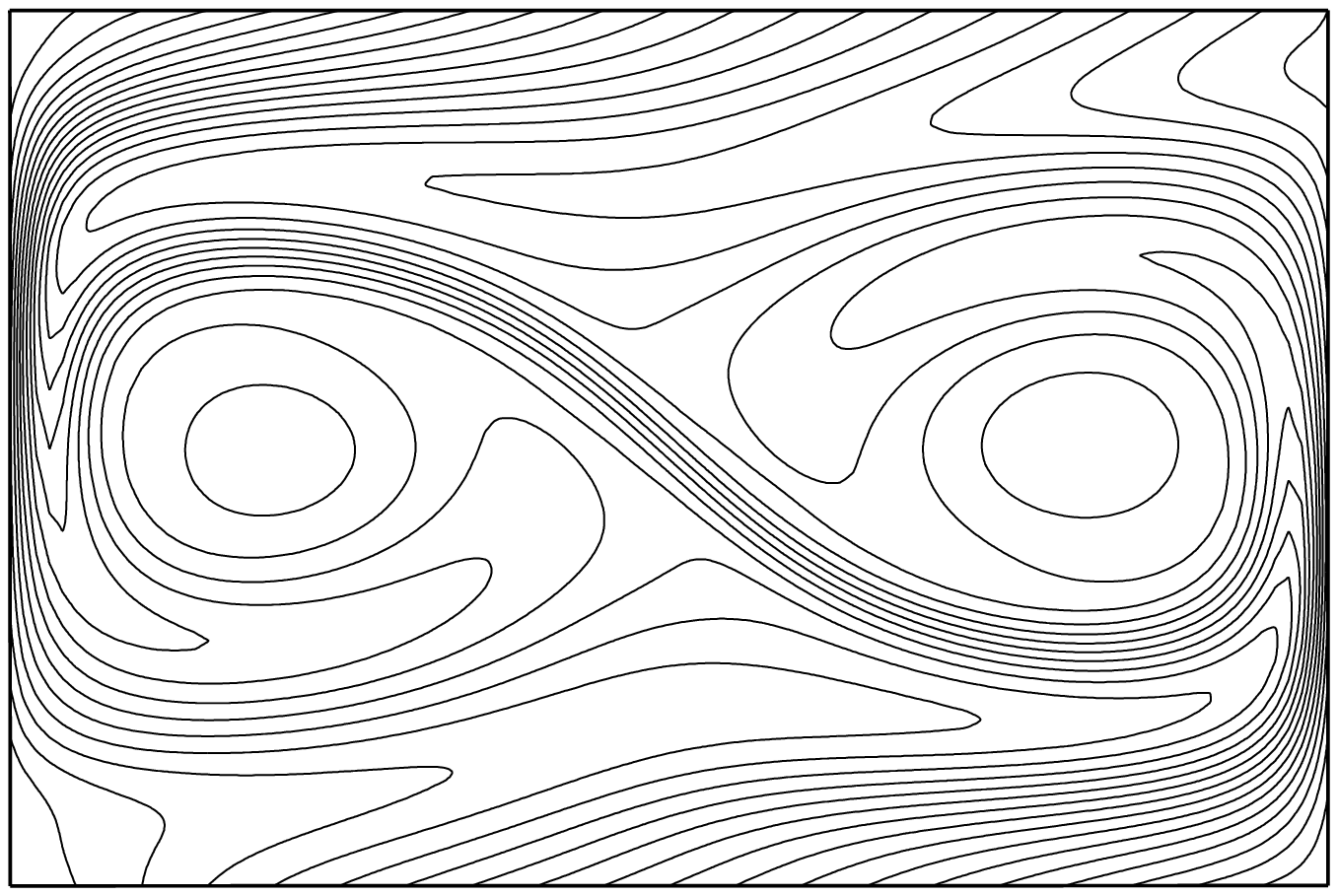}}}
\put(4,2){\makebox(0,0){
\includegraphics[width=6.5cm,angle=90.0]{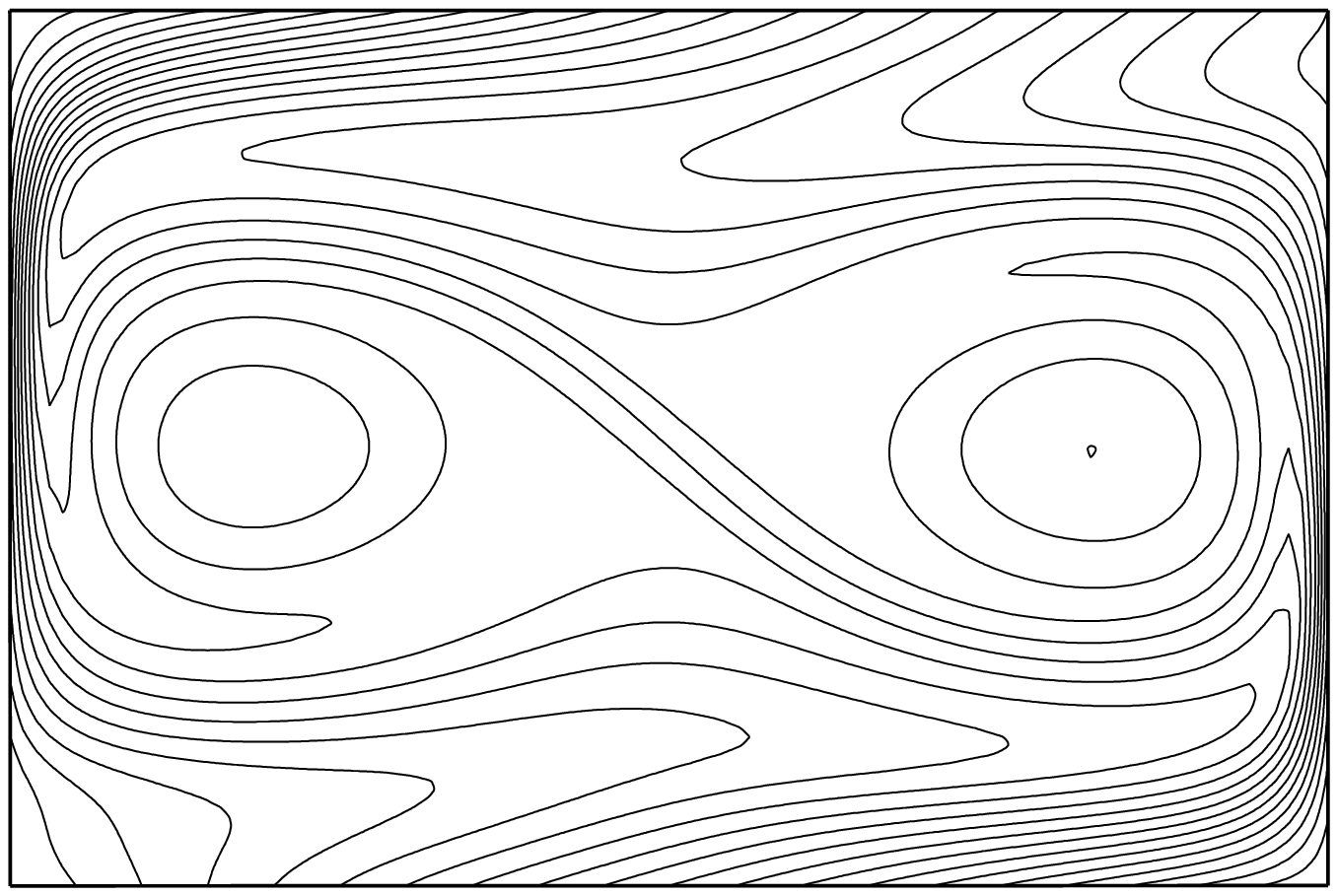}}}
\put(11,2){\makebox(0,0){
\includegraphics[width=6.5cm,angle=90.0]{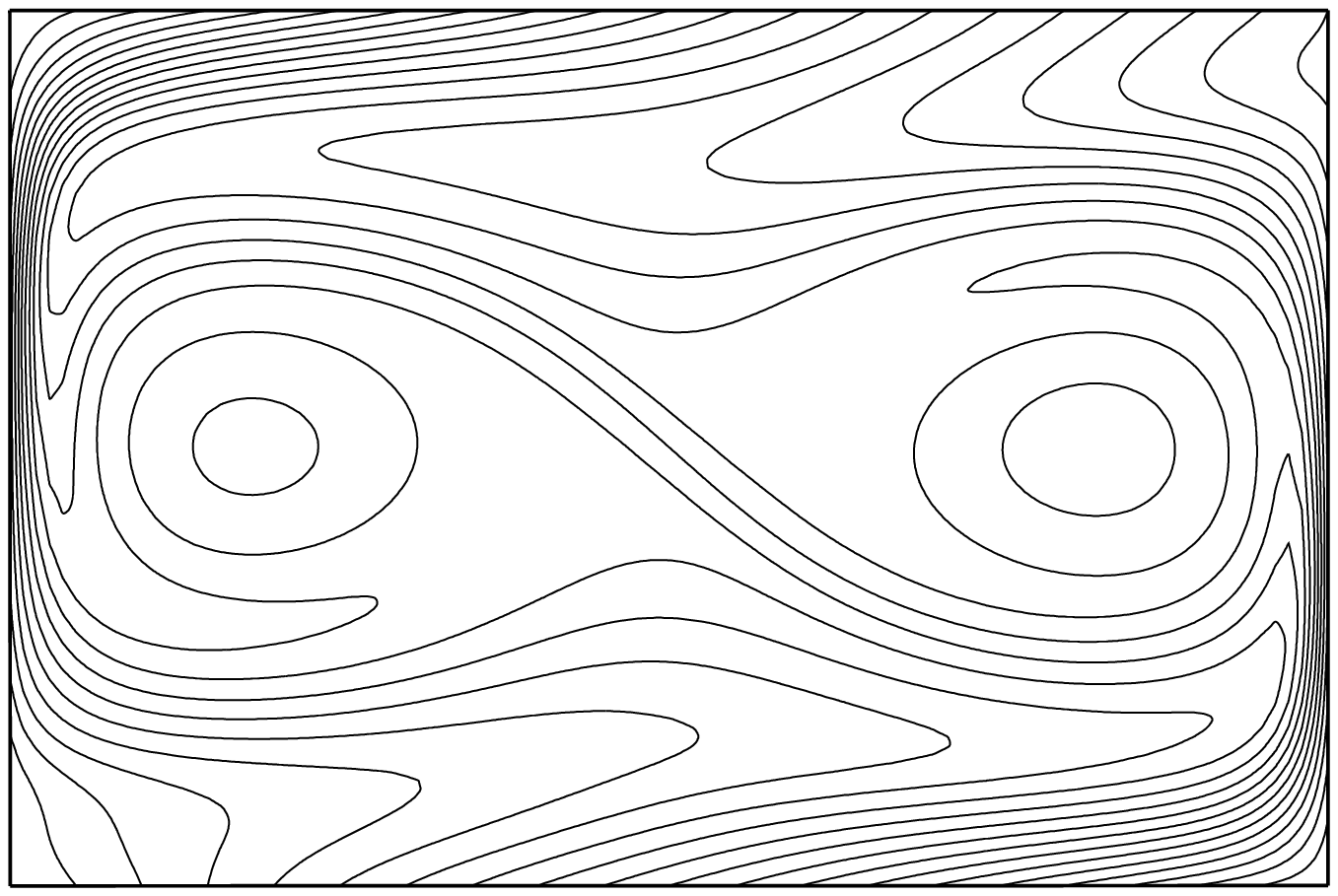}}}
\end{picture}}
\caption[Sequence 1]{% Figure 2:  
The stream function for $ m = 0.1 $ (a) and the corresponding time 
evolution of the magnetic field with $ R_M = 1000 $ (b-f) at times 
$ t = 1, \, 1.5, \, 2, \, 2.5, \, 3 $.}
\end{figure}

\begin{figure}[hbt] %[p]
\centerline{\unitlength1cm
\begin{picture}(15,8)
\put(4,6){\makebox(0,0){
\includegraphics[width=6.5cm,angle=90.0]{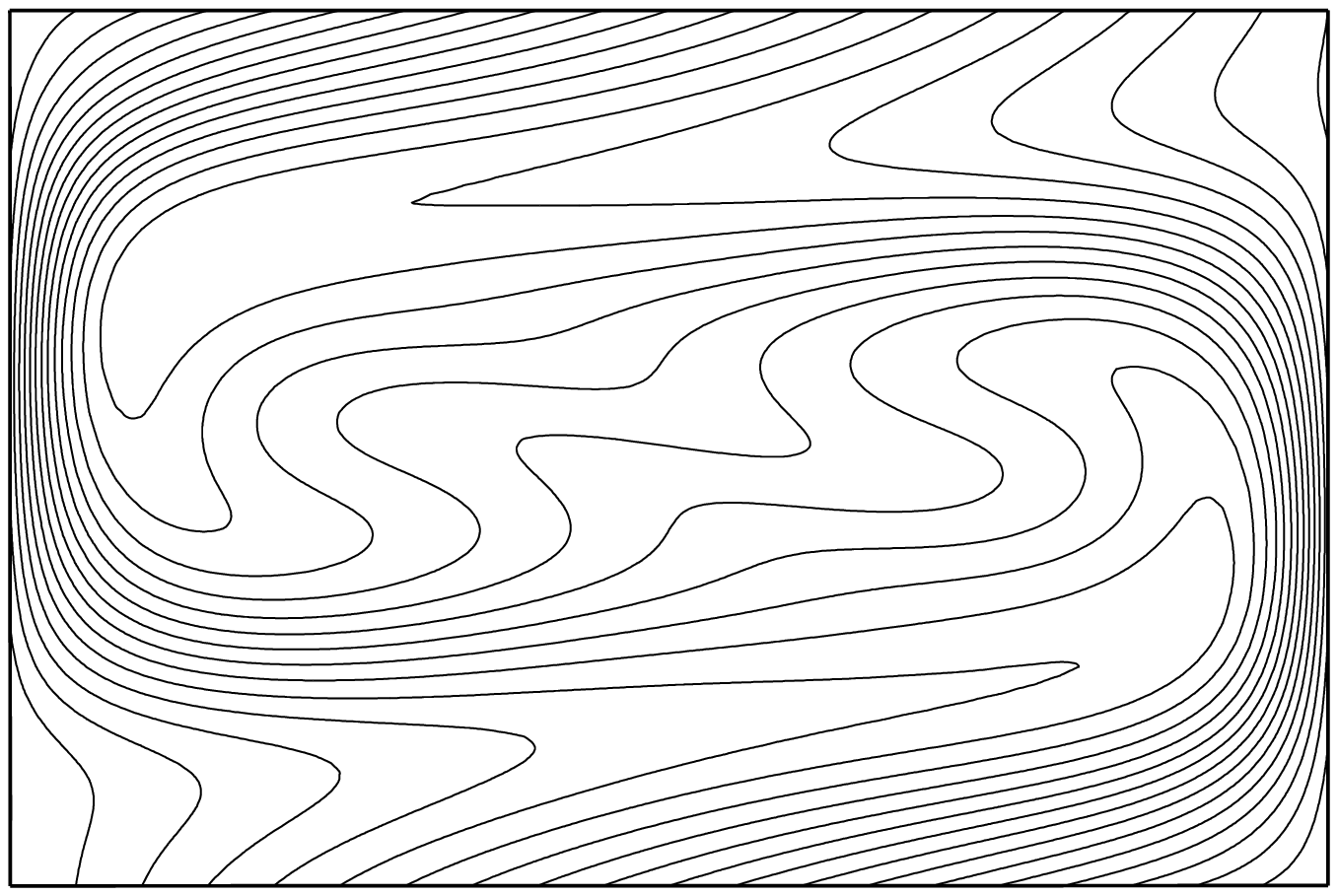}}}
\put(11,6){\makebox(0,0){
\includegraphics[width=6.5cm,angle=90.0]{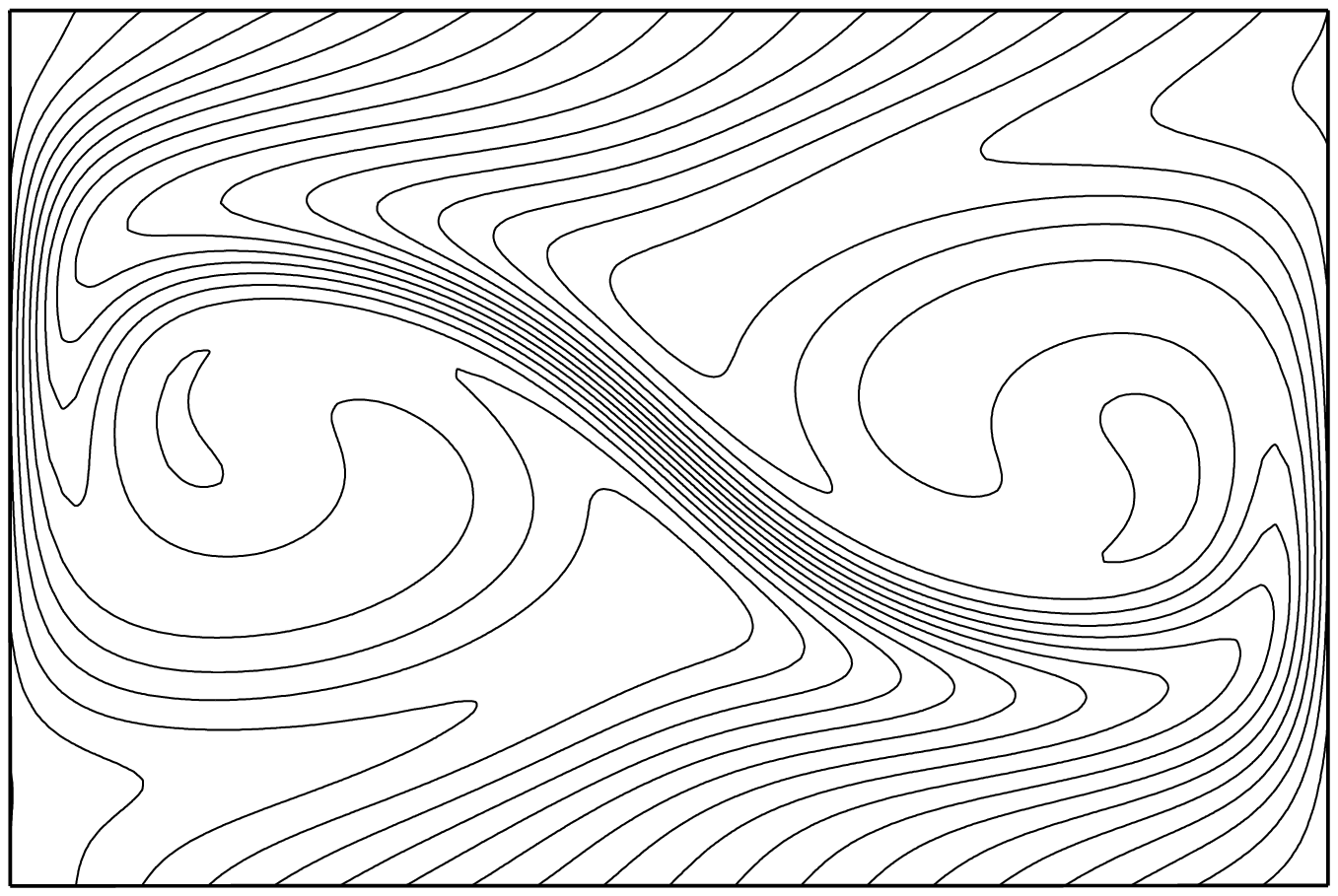}}}
\put(4,2){\makebox(0,0){
\includegraphics[width=6.5cm,angle=90.0]{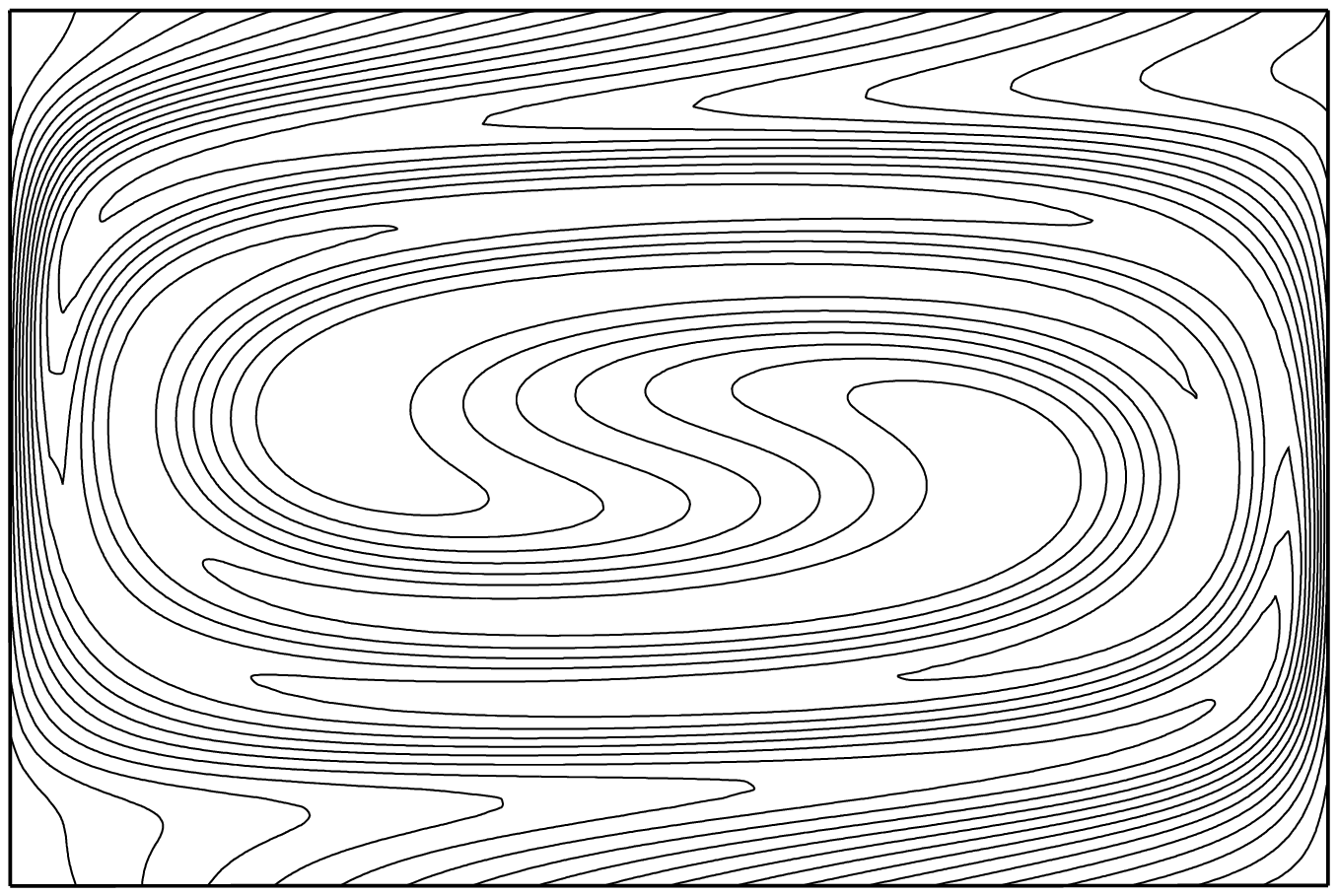}}}
\put(11,2){\makebox(0,0){
\includegraphics[width=6.5cm,angle=90.0]{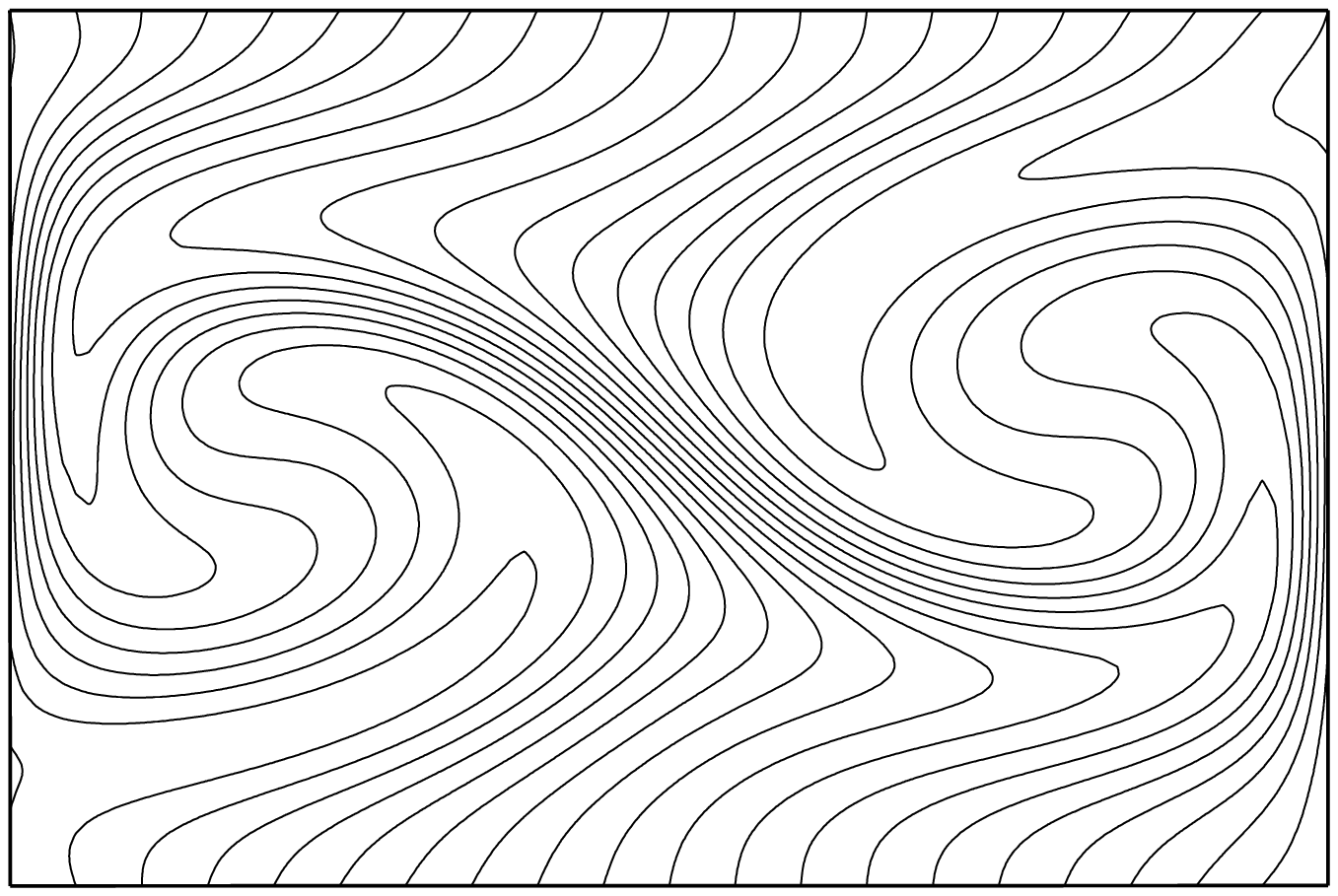}}}
\end{picture}}
%\vskip-2cm
\caption[Sequence 2]{% Figure 3:  
Magnetic field states for different magnetic 
Reynolds numbers and plasma velocity fields: (a) $ R_m = 200, \, m = 0.1 $, 
(b) $ R_m = 500, \, m = 0.3 $, (c) $ R_m = 800, \, m = 0.7 $ and 
(d) $ R_m = 1000, \, m = 0.1 $.}
\end{figure}

\begin{figure} %[p]
% ht = here, b = bottom, t = top, p = page of float
\label{Reynolds}
\centering
\includegraphics[width=13cm]{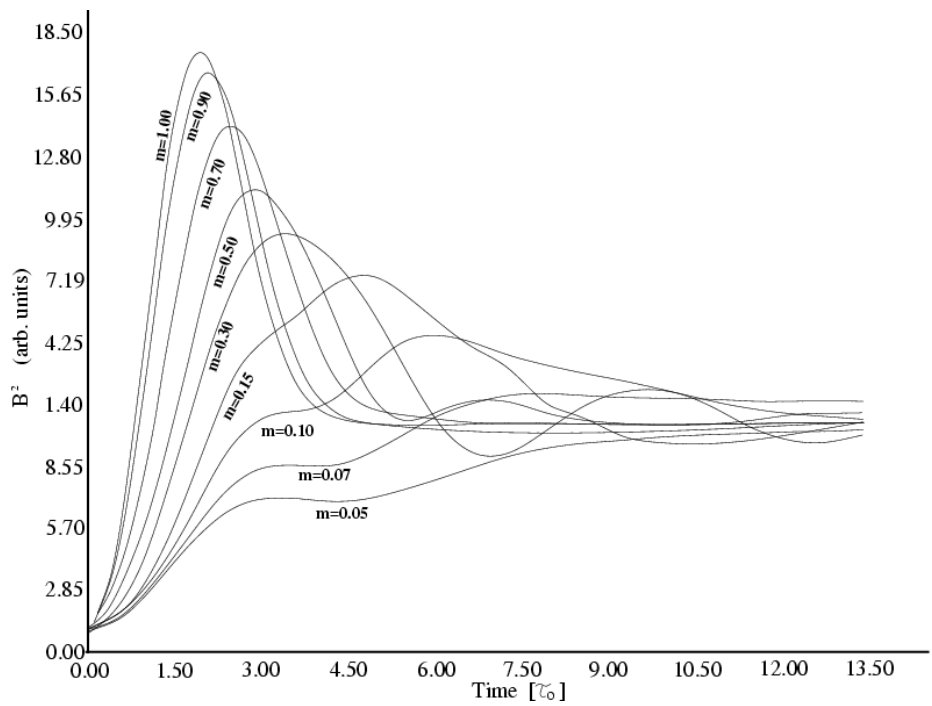}
\caption[Square of magnetic field]{% Figure 4:  
Square of magnetic field as a function of time for different $ m $ values.}
\end{figure}

\begin{figure} %[p]
% ht = here, b = bottom, t = top, p = page of float
\label{Stationary}
\centering{\includegraphics[width=11cm]{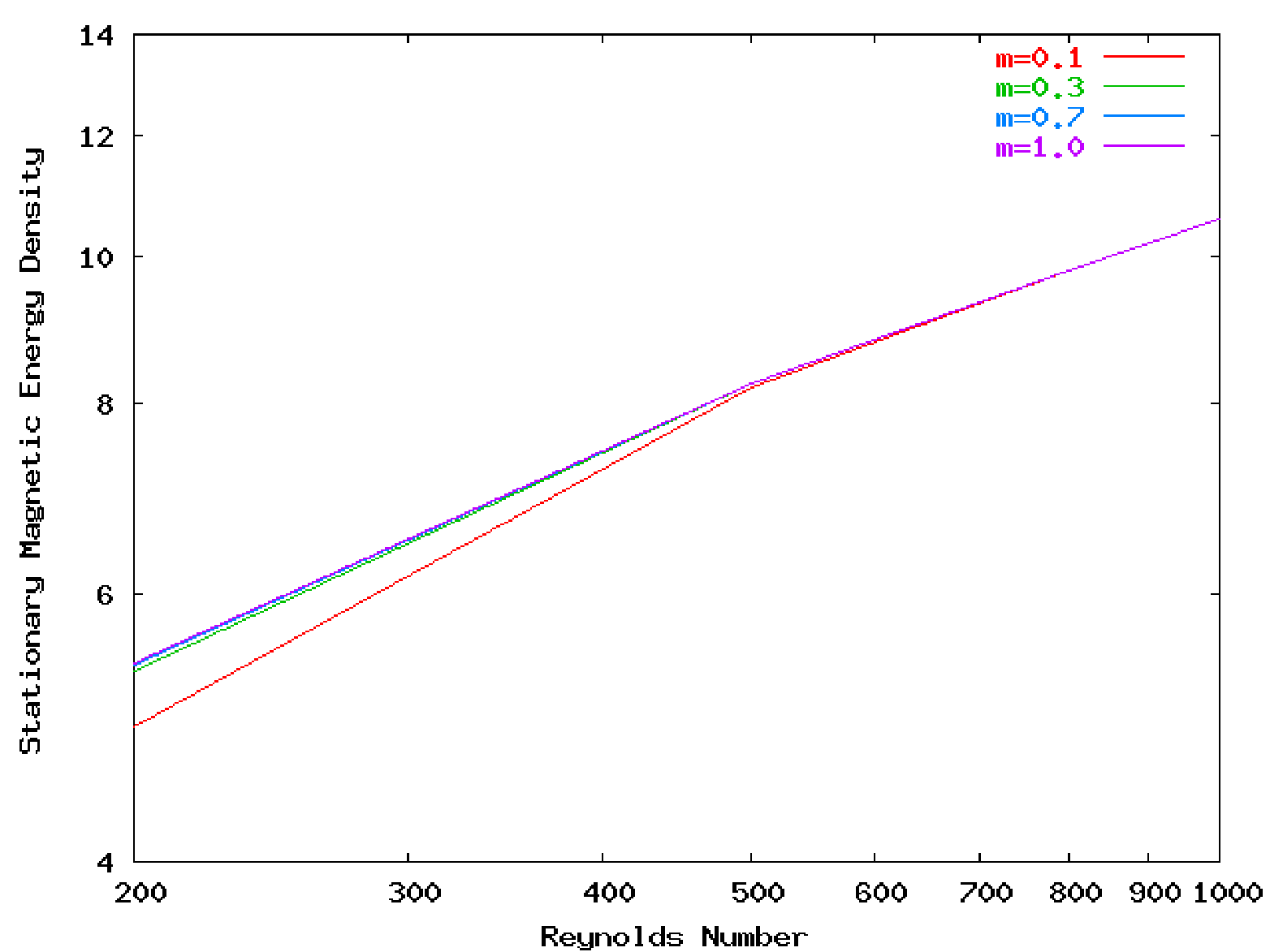}}
\centering{\includegraphics[width=11cm]{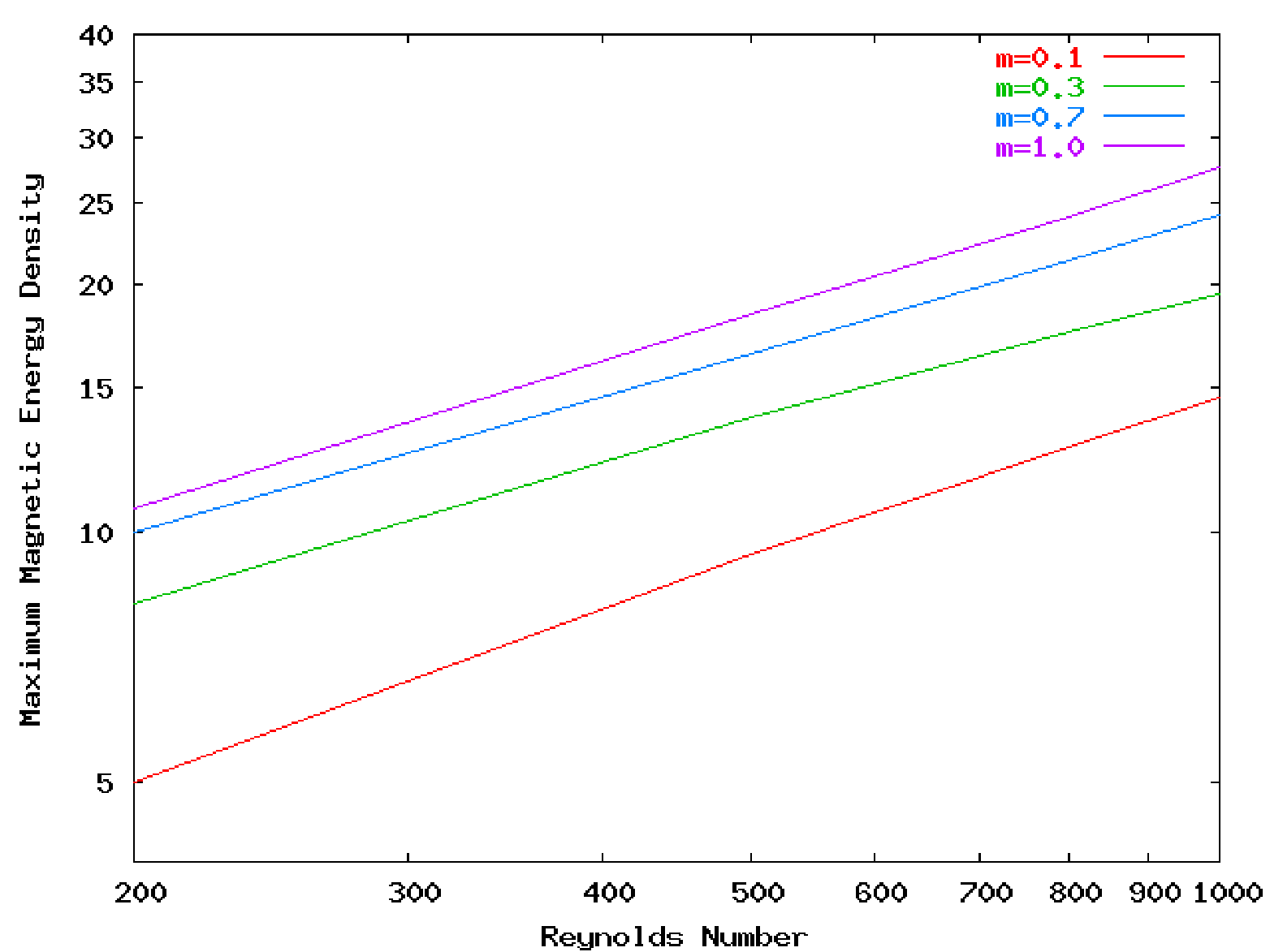}}
\caption[Stationary]{% Figure 5:  
Stationary (a) and maximum magnetic energy density (b) as function of 
the magnetic Reynolds number for different $ m $ values.}
\end{figure}


\begin{thebibliography}{99}

\bibitem{Carboni}
R.~Carboni, {\it Convective Plasmas}, M.~Sc.~Thesis, School of Physics, 
University of Costa Rica, 1994.    

\bibitem{Carboni2}
R.~Carboni, and F.~Frutos-Alfaro, 
{\it PCell: A 2D Program for Visualizing Convective Plasma Cells}, 
{Computing in Science and Engineering},  2 - 5,  July/August,  2004.

\bibitem{Chang}
T.~Chang, S.~W.~Y.~Tam and C.~Wu, {\it Complexity induced anisotropic 
bimodal intermittent turbulence in space plasmas}, {Physics of Plasmas}, 
11, 1287 - 1299, 2004.

\bibitem{Dolginov}
A.~Z.~Dolginov, {\it Magnetic field generation in celestial bodies}, 
{Physics Reports} 162: 337 - 415, 1988.

\bibitem{Elsasser}
W.~M.~Elsasser, {\it Hydromagnetic Dynamo Theory}, 
{Reviews of Modern Physics}, 28, 135 - 163, 1956.

\bibitem{Foukal}
P.~Foukal, {\it Solar Astrophysics}, Wiley \& Sons, New York, 1990.

\bibitem{Galloway} 
P.~A.~Galloway, M.~R.~E.~Proctor and N.~O.~Weiss, 
{\it Magnetic flux ropes and convection}, {Journal of Fluid Mechanichs}, 
87, 243 - 254, 1978.

\bibitem{Galloway2}
P.~A.~Galloway, and N.~O.~Weiss, 
{\it Convection and Magnetic Fields in Stars}, {Astrophysical Journal}, 
243, 120 - 130, 1981.

\bibitem{Minini}
P.~D.~Mininni, D.~O.~Gomez, and S.~M.~Mahajan, 
{\it Dynamo Action in Magnetohydrodynamics and Hall-Magnetohydrodynamics}, 
{Astrophysical Journal}, 587, 472 - 481, 2003.

\bibitem{Parker}
E.~N.~Parker, {\it Kinematical hydromagnetic theory and its applications 
to the low solar photosphere}, {Astrophysical Journal}, 138, 552 - 562, 1975.

\bibitem{Sturrock}
P.~A.~Sturrock, T.~Holzer, D.~M.~Mihalas, D. M., R.~K.~Ulrich, (Editors), 
{\it Physics of the Sun}, Reidel, Dordrecht,  1986.  

\bibitem{Valdivia}
J.~A.~Valdivia, A.~Klimas, D.~Vassiliadis, V.~Uritsky and J.~Takalo, 
{\it Self-organization in a current sheet model}, {Space Science Reviews}, 
107, 515 - 522, 2003.

\bibitem{Weiss}
N.~O.~Weiss, {\it The Expulsion of Magnetic Flux by Eddies}, 
{Proceedings of the Royal Society}, 293, 310 - 328, 1966.

\bibitem{Weiss2}
N.~O.~Weiss, {\it Magnetic Fields and Convection}, 
{Advanced Chemical Physics}, 32, 101 - 110, 1975.

\bibitem{Zegeling}
P.~A.~Zegeling, {\it On resistive MHD models with adaptive moving meshes}, 
{Journal of Scientific Computing}, Special ICIAM issue, 2005.

\bibitem{Zegeling2}
P.~A.~Zegeling and R.~Keppens, in {\it Adaptive Method of Lines} by 
A.~Vande~Wouwer, Ph.~Saucez, and W.~E.~Schiesser (Editors), 
Chapman \& Hall/CRC, 2001.

\end{thebibliography}
\end{document}